\documentclass[conference]{IEEEtran}
\IEEEoverridecommandlockouts
\usepackage{cite}
\usepackage{amsmath,amssymb,amsfonts}
\usepackage{algorithmic,array,multirow}
\usepackage{graphicx}
\usepackage{times}
\usepackage{url}
\usepackage{latexsym}

\usepackage{graphicx}
\usepackage{color}
\usepackage{array,multirow}
\usepackage{amssymb}
\usepackage[table]{xcolor}
\usepackage{subcaption}
\usepackage{wrapfig,lipsum,booktabs}
\usepackage{siunitx}
\usepackage{tikz}

\usepackage{booktabs,siunitx}
\def\BibTeX{{\rm B\kern-.05em{\sc i\kern-.025em b}\kern-.08em
    T\kern-.1667em\lower.7ex\hbox{E}\kern-.125emX}}

\newcommand{\mc}[2]{\multicolumn{#1}{#2}}
\usepackage{booktabs,siunitx}
\DeclareMathOperator*{\argmax}{argmax} 

\begin{document}

\title{Computational Pronunciation Analysis in Sung Utterances
\thanks{ED received funding from the European Union's Horizon 2020 research and innovation programme under the Marie Skłodowska-Curie grant agreement No.\ 765068.}
}

\author{\IEEEauthorblockN{1\textsuperscript{st} Emir Demirel}
\IEEEauthorblockA{\textit{Centre for Digital Music} \\
\textit{Queen Mary University of London}\\
London, UK \\
e.demirel@qmul.ac.uk}
\and
\IEEEauthorblockN{2\textsuperscript{nd} Sven Ahlb\"ack}
\IEEEauthorblockA{\textit{Doremir Music Research AB} \\
Stockholm, Sweden \\
sven.ahlback@doremir.com}
\and
\IEEEauthorblockN{3\textsuperscript{rd} Simon Dixon}
\IEEEauthorblockA{\textit{Centre for Digital Music} \\
\textit{Queen Mary University of London}\\
London, UK \\
s.e.dixon@qmul.ac.uk}
}

\maketitle

\tikz [remember picture, overlay] %
\node [shift={(34mm,22.5mm)}] at (current page.south west) %
[anchor=south west] %
{\includegraphics[width=7mm]{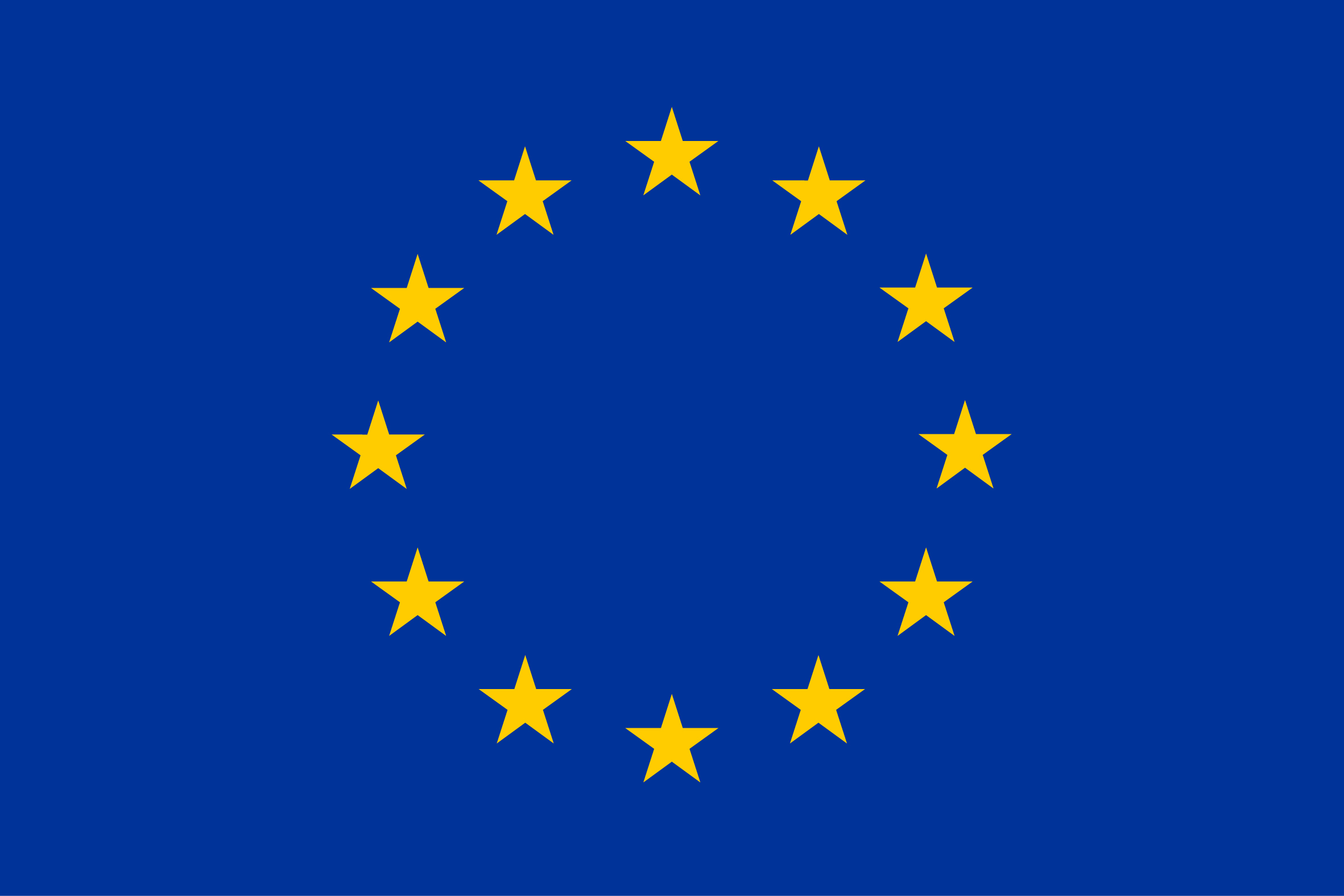}};

\begin{abstract}
Recent automatic lyrics transcription (ALT) approaches focus on building stronger acoustic models or in-domain language models, while the pronunciation aspect is seldom touched upon. This paper applies a novel computational analysis on the pronunciation variances in sung utterances and further proposes a new pronunciation model adapted for singing. The singing-adapted model is tested on multiple public datasets via word recognition experiments. It performs better than the standard speech dictionary in all settings reporting the best results on ALT in a capella recordings using n-gram language models. For reproducibility, we share the sentence-level annotations used in testing, providing a new benchmark evaluation set for ALT.
\end{abstract}


\section{Introduction}

The articulation of words during singing is influenced by the melodic line causing temporal variations in duration and the acoustic properties of the signal like pitch, timbre and loudness. Singers may even add an extra formant on top of the ones that characterize vowels, namely the singer's formant \cite{sundberg1990science}, increasing the perceived loudness of the voice. Consequently, these articulations during singing may alter the ways that words are pronounced and how they are perceived, thus affecting overall intelligibility \cite{smith1980}. Similarly, the performance of ALT systems that attempt to automatically recognize words from singing voice also gets affected by these altered pronunciations and variations in the acoustic properties. While only a little focus has been drawn to computationally model the pronunciation variances in singing performances, Gupta et al.\cite{gupta2018automatic} proposed to use a vowel-extended version of a standard lexicon with regards to the longer vowels in sung utterances and observed considerable improvement in word recognition.

In this study, we aim to shed a light on the pronunciation differences in sung utterances compared to speech by conducting a novel quantitative analysis on the phoneme level, identifying a number of systematic cases. Furthermore, we propose a new lexicon adaptation method for modelling of singing, and evaluate its effectiveness through word recognition rates over a number of open-source data sets. Additionally, we test the findings of the phonetic analysis through an error analysis. For reproducibility, we share the annotations publicly.

This paper begins by establishing a contextual ground to understand how the pronunciation models are employed in common hybrid-ASR frameworks. In Section 3, the details of the phonetic analysis are presented. Then, a method is proposed for extending a standard pronunciation dictionary for singing with respect to the observations of the analyses in Section 3. Section 4 explains the ALT setup for the recognition experiments. Section 5 provides the error analysis in terms of word and character error rates. 
\newcommand{\STAB}[1]{\begin{tabular}{@{}c@{}}#1\end{tabular}}    

\section{Related Work}

The task of ASR can be summarized as finding the most probable word sequence, $\widehat{\mathbf{w}}$, given a sequence of acoustic observations, $\mathbf{X}$, which can be expressed using the following formula:

\begin{equation}
    \widehat{\mathbf{w}} = \argmax_\mathbf{w} P(\mathbf{w}) \sum_{\mathbf{Q} \in \mathbf{Q}_w}, P(\mathbf{X}|\mathbf{Q}) P(\mathbf{Q}|\mathbf{w})
\end{equation}

\noindent where $\mathbf{Q}$ is a sequence of phonemes  and $\mathbf{Q}_w$ is the set of all possible state sequences that correspond to the word sequence $\mathbf{w}$, as defined by a lexicon (i.e.\ pronunciation dictionary) \cite{gales2008application}. Then, $P(\mathbf{Q}|\mathbf{w})$ gives the probability of observing a certain phoneme sequence belonging to a word.

The speech recognition framework we use, Kaldi \cite{povey2011kaldi}, constructs the decoding graph via Weighted Finite State Transducers (WFSTs) \cite{mohri2008speech}. The final decoding graph $\mathit{HCLG}$ is a composition of multiple finite-state transducers:

\begin{equation}
    \mathit{HCLG} = H \circ C \circ L \circ G,
    \label{eqn:hclg}
\end{equation}

\noindent where $\circ$ is the operation of graph composition for finite-state transducers (FST), and each element in Equation \ref{eqn:hclg} represents a FST \cite{mohri2008speech}. In summary, the phoneme posteriors are obtained via the acoustic model in $H$, and the HMM phone states are converted/relabeled to context-dependent `triphone' states via $C$. Finally, the operation $L \circ G$ pairs any word string $w$ in a pronunciation lexicon to its corresponding pronunciation  $q^{w}$.

While it is common to use the standard CMUSphinx English Pronunciation Dictionary \cite{weide1998cmu} in lyrics transcription\cite{dabike2019automatic,demirel2020}, a vowel-extended version has also been shown to be beneficial for word recognition\cite{gupta2018automatic}. In this paper, we aim to study the pronunciation variances in singing through a confusion analysis, a procedure similar to the ones presented in \cite{morales2007,yilmaz2014}. Based on profound phonemic confusions, we create an extended version of the standard CMU dictionary by adding alternative word pronunciations for singing, which has been shown to be an effective method in ASR\cite{lexica2000}.

\section{Pronunciation Analysis}\label{sec:analysis}

The phonetic analysis is based on the confusions between orthographic transcriptions, $\widehat{\mathbf{Q}}$, produced by a pretrained ALT model that uses a pronunciation dictionary for speech (CMUSphinx \cite{weide1998cmu}) and the human phoneme annotations, $\mathbf{Q}$, on the singing performances chosen for analysis. We use the \textit{NUS Sung and Spoken Lyrics Corpus} \cite{duan2013nus} due to the availability of phoneme-level annotations, and choose native English speaking (with North American accent) singers \textit{f01, f02, m09, m11} for analysis. We limit our analysis to these singers in order to minimize the influence of non-native accents.

Initially, the word transcriptions $\widehat{\mathbf{W}}$ are extracted and decomposed into their phonemic representations $\widehat{\mathbf{Q}}$ by decomposing the lexicon transducer $L$ from the decoding graph $\mathit{HCLG}$. To get the phoneme confidences, we align $\widehat{\mathbf{Q}}$ with their corresponding manually annotated phoneme sequences. During this alignment, we take the following steps:

\begin{enumerate}
    \item We compute the alignment score matrix $\mathbf{D}$ by performing Levenshtein alignment, \textit{lev}, between the phoneme tokens $q$ of the predictions $\widehat{\mathbf{Q}}_M$ and the ground truth $\mathbf{Q}_N$:
    \begin{equation}
        \mathbf{D}_{M \times N} = \mathit{lev}({\widehat{\mathbf{Q}}_M},\mathbf{Q}_N),
    \end{equation}
    and find the best alignment path, $\mathbf{A}_{2 \times K}$ through reverse tracing to find the path with the lowest pairwise gap cost: 
    \begin{equation}
    \mathbf{A}_{2 \times K} = 
    \begin{pmatrix}
    \dots & q_{k-1} & q_{k} & q_{k+1} & \dots\\
    \dots & \widehat{q}_{k-1} & \widehat{q}_{k} & \widehat{q}_{k+1} & \dots
    \end{pmatrix}.
    \end{equation}
    $\mathbf{A}$ can be interpreted as a sequence of phoneme pairs.
    \item  There are three operations defined on these phoneme pairs to match $\widehat{\mathbf{Q}}_M$ to $\mathbf{Q}_N$: insertions ($I$), substitutions ($S$) and deletions ($D$). These operations are represented in $\mathbf{A}$ with the symbol $\epsilon$. An alignment instance  $ a_k = \begin{pmatrix} \epsilon  \\ \widehat{q}^*_k \end{pmatrix}$ is a deletion and the opposite case would be an insertion.
    \item Let the number of correctly matching pairs in $\mathbf{A}$ be $C$, then the confidence score per phoneme type, $c_q$, can be retrieved as:
    \begin{align}
    \label{eqn:c_phi}
        c_q = \frac{\sum_{i}^T C_{q,i} - (S_{q,i} + I_{q,i }+ D_{q,i})}{\sum_i^T  C_{q,i} + S_{q,i} + I_{q,i }+ D_{q,i}}, \nonumber \\ q \in \Omega_E,
    \end{align}
    where $T$ is the number of utterances in the analysis set and $\Omega_E$ is the English phoneme set used in our analysis. The denominator is necessary to normalize with respect to the total number of pairs in $\mathbf{A}$, since the phonemes in $\Omega_E$ are not necessarily represented equally in the analysis data set.
\end{enumerate}

\begin{table}[ht!]
\centering
\scalebox{0.82}{
\begin{tabular}{|p{2.1cm} | p{0.5cm} | p{1cm}|  p{1.4cm}|}
\hline
\textbf{Vowels} & $q$ & $c_{q} (R)$ & \hfil${\Phi'_N}$\\\hline
\multirow{5}{*}{Short Vowels} & \cellcolor{darkgray!6}\scriptsize{AE} & \cellcolor{darkgray!6}\scriptsize{-0.42 (38)}& \cellcolor{darkgray!6}\scriptsize{AH, EH, AA}\\ \cline{2-4}
                                                                 & \cellcolor{darkgray!6}\scriptsize{AH} & \cellcolor{darkgray!6}\scriptsize{0.17 (33)}& \cellcolor{darkgray!6}\scriptsize{AA,EH,OW}\\\cline{2-4}
                                                                 & \scriptsize{EH} & \scriptsize{0.3 (32)}&\scriptsize{AH,AE,IH}\\\cline{2-4}
                                                                & \scriptsize{IH} & \scriptsize{0.48 (26)}&\scriptsize{IY,AH,EY}\\\cline{2-4}
                                                                 & \cellcolor{darkgray!6}\scriptsize{UH} & \cellcolor{darkgray!6}\scriptsize{0 (36)}&\cellcolor{darkgray!6}\scriptsize{AO,UW,AH}\\\cline{1-4}
     \multirow{6}{*}{Long Vowels} & \scriptsize{AA} & \scriptsize{0.5 (24)}&\scriptsize{AO,AW,AE}\\\cline{2-4}
                             & \cellcolor{darkgray!6}\scriptsize{AO} & \cellcolor{darkgray!6}\scriptsize{0.06 (35)}&\cellcolor{darkgray!6}\scriptsize{AA,AH,OW}\\\cline{2-4}
                             & \cellcolor{darkgray!6}\scriptsize{ER} & \cellcolor{darkgray!6}\scriptsize{0.36 (31)}&\cellcolor{darkgray!6}\scriptsize{AH,OW,EH}\\\cline{2-4}
                             & \scriptsize{IY} & \scriptsize{0.87 (6)}&\scriptsize{EY,IH,EH}\\\cline{2-4}
                             & \scriptsize{UW} & \scriptsize{0.88 (4)}&\scriptsize{OW,AH,UH}\\\cline{1-4}
     \multirow{4}{*}{Diphthongs} & \scriptsize{AY} & \scriptsize{0.86 (8)}&\scriptsize{AA,AH,EH}\\\cline{2-4}
                             & \scriptsize{AW} & \scriptsize{0.71 (18)}&\scriptsize{AA,AH}\\\cline{2-4}
                             & \scriptsize{EY} & \scriptsize{0.87 (7)}&\scriptsize{IY,AY,EH}\\\cline{2-4}
                             & \scriptsize{OW} & \scriptsize{0.76
                (17)}&\scriptsize{AO,AA,AH}\\\cline{2-4}
                             & \scriptsize{OY} & \scriptsize{0.4 (28)}&\scriptsize{OW,AO,AY} \\
\hline
\end{tabular}
}
\caption{Results of the phonetic analysis (vowels)}
\label{table:vowels}
\end{table}

Tables \ref{table:vowels} and \ref{table:consonants} show the results of the phoneme confusion analysis for vowels and consonants respectively. The first two columns from the left are the list of English phoneme categories and types\footnote{We use the standard 39-phoneme set of the CMU dictionary.}. In the middle column, the confidence scores and their confidence rankings $R$ are provided. By definition in Equation \ref{eqn:c_phi}, $-1 \leq c_q \leq 1$, hence we did not further normalize this value. According to Equation \ref{eqn:c_phi}, $c_q < 0.25$ means that there are less true positives than the sum of false negatives and positives in per phoneme type predictions, i.e. in most cases, $q$ is predicted incorrectly. The phonemes in the rightmost column,  ${\Phi'}$, are determined according to the most frequent instances of substitutions.

\begin{table}[h!]
\centering
\scalebox{0.82}{
\begin{tabular}{|p{2.1cm} | p{0.5cm} | p{1cm}|  p{1.4cm}|}
\hline
{\textbf{Consonants}} & $q$ & $c_{q} (R)$ & \hfil${\Phi'_N}$\\\hline
  \multirow{6}{*}{Plosives} & \scriptsize{B} & \scriptsize{0.77 (16)}&\scriptsize{D,P,W}\\ \cline{2-4}
                                                                & \cellcolor{darkgray!6}\scriptsize{D} & \cellcolor{darkgray!6}\scriptsize{0.16 (34)}&\cellcolor{darkgray!6}\scriptsize{T,N,JH}\\\cline{2-4}
                                                                & \scriptsize{G} & \scriptsize{0.77 (15)}&\scriptsize{NG,K}\\\cline{2-4}
                                                                & \scriptsize{K} & \scriptsize{0.85 (15)}&\scriptsize{G,HH}\\\cline{2-4}
                                                                & \scriptsize{P} & \scriptsize{0.78 (14)}&\scriptsize{B,M,F}\\\cline{2-4}
                                                                & \cellcolor{darkgray!6}\scriptsize{T} & \cellcolor{darkgray!6}\scriptsize{0.37 (29)}&\cellcolor{darkgray!6}\scriptsize{D,S,CH}\\\cline{1-4}
      \multirow{2}{*}{Affricates}                            & \scriptsize{CH} & \scriptsize{0.79 (13)}&\scriptsize{JH,SH,T}\\\cline{2-4}
                                                                & \scriptsize{JH} & \scriptsize{0.88 (5)}&\scriptsize{CH,S,Y}\\\cline{1-4}
      \multirow{3}{*}{Nasals}                               & \scriptsize{M }& \scriptsize{0.93 (2)}&\scriptsize{N,NG}\\\cline{2-4}
                                                                & \scriptsize{N} & \scriptsize{0.85 (12)}&\scriptsize{M,NG,D}\\\cline{2-4}
                                                                & \scriptsize{NG }& \scriptsize{0.85 (9)} &\scriptsize{N,M,T}\\\cline{1-4}
     \multirow{6}{*}{Fricatives}                              & \cellcolor{darkgray!6}\scriptsize{DH} & \cellcolor{darkgray!6}\scriptsize{0.36 (30)}&\cellcolor{darkgray!6}\scriptsize{TH,D,N}\\\cline{2-4}
                                                                & \scriptsize{F }& \scriptsize{0.91 (3)}&\scriptsize{V,P,TH}\\\cline{2-4}
                                                                & \scriptsize{HH} & \scriptsize{0.70 (19)}&\scriptsize{DH,W,Y}\\\cline{2-4}
                                                               & \scriptsize{S} & \scriptsize{0.95 (1)}&\scriptsize{Z,TH,T}\\\cline{2-4}
                                                                & \scriptsize{SH} & \scriptsize{0.85 (10)}&\scriptsize{CH,S,Z}\\\cline{2-4}
                                                                & \scriptsize{TH} & \scriptsize{0.57 (21)}&\scriptsize{S,T,DH}\\\cline{2-4}
                                                               & \scriptsize{V} & \scriptsize{0.56 (22)}&\scriptsize{F,R,DH}\\\cline{2-4}                                           
                                                                & \cellcolor{darkgray!6}\scriptsize{Z} & \cellcolor{darkgray!6}\scriptsize{-0.05 (37)}&\cellcolor{darkgray!6}\scriptsize{S,T}\\\cline{2-4} 
                                                                & \scriptsize{ZH} &\scriptsize{N/A} & \scriptsize{N/A} \\\cline{1-4} 
      \multirow{4}{*}{Approximants*}                         & \scriptsize{L }& \scriptsize{0.44 (27)}&\scriptsize{AA,OW,AH}\\\cline{2-4}
                                                                & \scriptsize{R} & \scriptsize{0.48 (25)}&\scriptsize{AA,AH,IH}\\\cline{2-4}
                                                                & \scriptsize{W} & \scriptsize{0.66 (20)} & \scriptsize{AA,OW,V}\\\cline{2-4}
                                                                & \scriptsize{Y} & \scriptsize{0.55 (23)}&\scriptsize{IH, AH, IY}\\
\hline
\end{tabular}
}
\caption{Results of the phonetic analysis (consonants)}
 \label{table:consonants}
\end{table}

It can be seen from the aforementioned tables that it is mostly the vowels that have the lowest confidences. Among vowels, \textit{diphthongs} are more accurately predicted than \textit{short vowels} and \textit{long vowels}. The phoneme `\textit{AE}' has the lowest $c_q$ and is generally associated with the \textit{schwa} sound in phonetics \cite{silverman2011schwa}. This very low $c_q$ is not surprising as it is often pronounced weakly and is one of the most frequently occurring vowel sounds in the English language \cite{roach2004british}.

For consonants, $c_q$ has higher values in general, but does not seem to be very consistent per phoneme category. The plosives `\textit{D}' and `\textit{T}' are severely confused indicating a systematic error, similarly for the other phonemes `\textit{DH}', `\textit{TH}' and `\textit{Z}'. On the other hand, plosives `\textit{B,G,K,P}' have rather high confidences. This is not extremely surprising as it has been mentioned in the literature that singers may utilize such phonemes to utter strong note offsets during melody construction\cite{bauer2002}. According to this observation, the singers in our analysis data did not seem to omit `\textit{B,G,K,P}' sounds. Though $\Phi'$ in different parent phoneme categories is not considered in Table \ref{table:consonants}, we observed systematic confusions in \textit{approximants} with vowels. This might be an indication of either systematic misalignment errors due to longer vowels, or omitted vowels for fluency during melody construction. In addition to the \textit{short vowels}, the `\textit{HH}' and `\textit{Y}' sounds are inserted the most to the predictions compared to the manual human annotations.

\begin{figure}
 \centering
 \includegraphics[clip,width=0.45\textwidth,height=5cm]{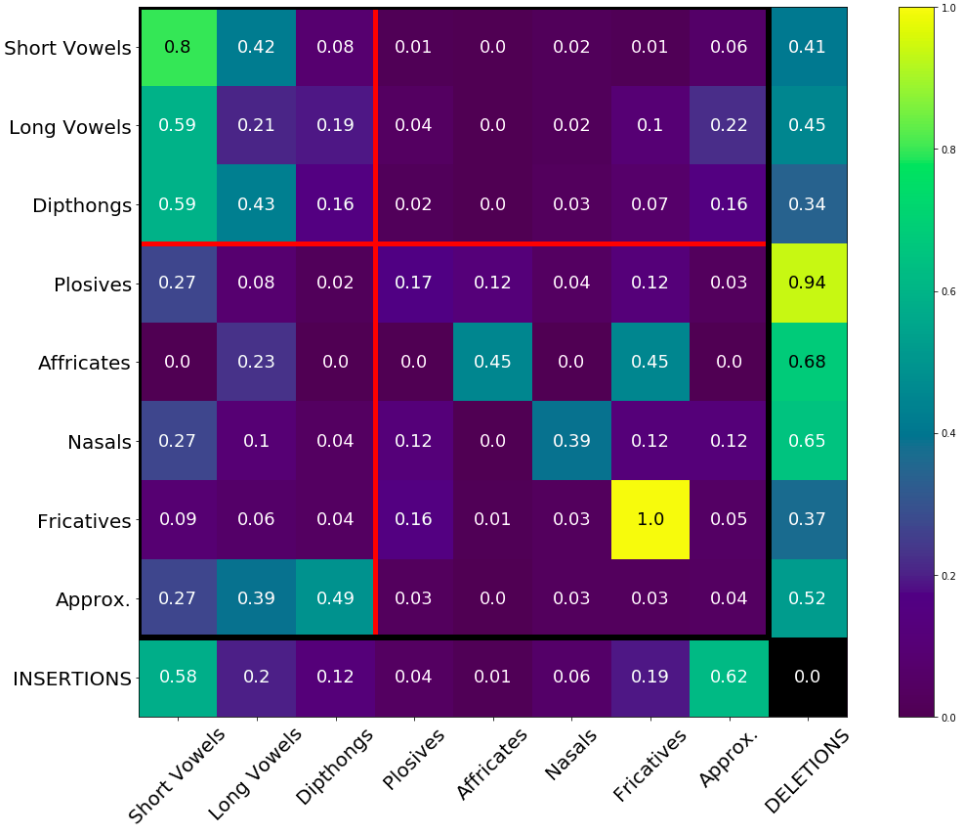}%
 \caption{Confusion matrix w.r.t.\ phoneme categories in Tables \ref{table:vowels} and \ref{table:consonants}. The red lines highlight the within-class regions for vowels and consonants. The numbers in cells are normalized values. The labels on the horizontal and the vertical axes represent the ground-truth and predictions respectively.}
 \label{fig:cm_plot}
\end{figure}

In Figure \ref{fig:cm_plot}, we show the phoneme confusion matrix summarized with respect to phonetic categories. We discard $C_q$ for calculating the confusions and sum only $S$, $I$ and $D$ values for each phonetic category. Therefore the diagonal axis does not represent self-confidences. Instead it represents the domestic confusions within each phonetic category. Phonetic-category-wise normalization is applied based on unit sum. These normalization steps are crucial to get confusion values independent of the number of occurrences. Insertions and deletions for each category are also included in the figure. The concentration of high confusion rates can be observed for vowels (top left). Short vowels are mostly confused with \textit{short vowels}. The annotated longer vowels are not necessarily represented in the standard speech lexicon, thus causing the system to assign a higher likelihood for the short vowels when making word predictions.  Note the high number of deleted \textit{plosives} signaling them being omitted from pronunciations during singing. Overall, a high frequency of deletions is observed. In addition to alignment errors, one possible cause for this could be the word liaisons being annotated as single phonemes in human annotations whereas the ALT system would predict such instances as separate phonemes. For example, in `\textit{DREAM MAKER}', `\textit{M}' is annotated once in the corresponding $\mathbf{Q}$, but detected twice in $\widehat{\mathbf{Q}}$.

\section{Extending the Lexicon}\label{sec:lexicon_ext}

In this section, we propose a pronunciation model for sung utterances based on the observations of the previous step. We extend the standard pronunciation model for speech through generating alternative pronunciations for singing.

\begin{figure}
\centering
 \includegraphics[clip,width=0.45\textwidth,height=2.7cm]{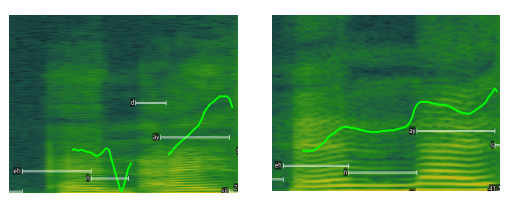}%
 \caption{An example of an omitted plosive in singing. $W$ = `AND I ' ; $Q^{\mathit{read}}$ = `AE N \textbf{D} AY' (left) ; $Q^{\mathit{sing}}$ = `EH N AY'. The gray horizontal lines show the temporal phoneme regions and the bright green curves are the pitch tracks extracted using pYIN \cite{mauch2014pyin}. }
 \label{fig:omitplosive}
\end{figure}

It is not seldom that in singing, performers may omit some consonants at the endings of words. This phenomenon can be explained as a
stylistic convention that singers exhibit in their performances in order to maintain the sonority of their singing \cite{sundberg1990science}, or it could as well be a microphone technique to avoid unpleasant pops. The analysis in Section \ref{sec:analysis} suggests that this occurs most likely for \textit{plosives} as the phoneme category with highest number of deletions. An example of an omitted \textit{plosive} is illustrated in Figure \ref{fig:omitplosive}\footnote{The analysis is performed on Sonic Visualizer software \cite{SonicVisualiser}.}. The spectrogram segments in Figure \ref{fig:omitplosive} show the same  words uttered as speech (left) and singing (right) by the same performer. According to the human annotators, the phoneme `\textit{D}', is not present during singing. This can also be seen from the discrepancies in the spectrogram and the undisturbed pitch curve in the singing segment\footnote{According to the empirical study in \cite{raymond2015}, pitch and phoneme perception are found to be cognitively correlated processes. Hence, we have chosen explicitly to show the pitch tracks. }. We add alternative pronunciations to such words ending with consonants \textit{D, T, DH} \% \textit{Z} by removing their last phoneme in the corresponding $q^{w_l}$. We have chosen these consonants due to them having the lowest confidences according to the analysis in Section \ref{sec:analysis}
.

It is noted in  \cite{kruspe2016bootstrapping} that longer vowels in singing may potentially cause alignment errors, consequently affecting the training and thus the recognition performance. Gupta et al.\ \cite{gupta2018automatic} proposed to extend the occurrences of vowels in each word in the lexicon for modeling longer vowels. Through representing longer vowels as consecutive repeated phonemes a better alignment in singing performances can be achieved, and hence a potential improvement in WER. In this study, we also apply a similar strategy when extending the lexicon. For instance, consider the word \textit{OCEANS} with its phonemic representation \textit{OW SH AH N Z} in the lexicon. We extend the occurrence of each vowel for up to 2 times, for example: \textit{OW OW SH AH N Z}, \textit{OW SH AH AH N Z} instead of 4 times (as in \cite{gupta2018automatic}) so that a smaller transducer is generated, for efficient decoding.

The final version of the singing-adapted lexicon is constructed by combining the two approaches mentioned in this section. The goal of this is to create a model that is generalizable to common pronunciation variances observed in Section 3.

\section{Experimental Setup}

We evaluate the effectiveness of the proposed singing-adapted lexicon with respect to word recognition rate via ALT experiments. We compare its performance in terms of word and character error rates (WER) with a model trained on the standard CMU English pronunciation dictionary.

For training, we utilize the train split of the \textit{DAMP} data set used by Demirel et al.\ \cite{demirel2020} which consists of approximately 150 hours of monophonic singing recordings of English language pop songs recorded in a Karaoke setting with a non-negligible proportion of noise. There are performers from 30 different countries in the data set, hence allowing a powerful acoustic model to generalize the accentual variations. 

\begin{table}[ht]
\centering
\scalebox{0.82}{
\begin{tabular}{ c !{\vrule width 2pt} c|  c | c | c }
  & Char. & Words & Sentences & Recordings    \\
  \hline
NUS\_read & 21935 & 5788 & 781 & 32 \\
NUS\_sing & 21935 & 5788 & 1029 & 32 \\
DAMP\_test & 17609 & 4626 & 479 & 70
\end{tabular}}
\caption{Statistics of evaluation sets}
\label{table:data}
\end{table}

For testing the lexicons, we have trained the lyrics transcriber using the pipeline in \cite{demirel2020} from the beginning at each experimental iteration. The transcriber in \cite{demirel2020} is based on a hybrid-ASR framework where the acoustic model consists of neural networks trained on lattice-free maximum mutual information (LF-MMI) setting\cite{povey2016purely}.  The neural network consists of stacks of 2D fully convolutional and factorized time-delay layers \cite{povey2018semi} with a self-attention layer added on top. At the input of the network, we extract 40-band filterbank features obtained with a hop size of 10ms and frame length of 20ms. To perform singer-adaptive training, we combine filterbank features with iVectors \cite{saon2013speaker}. The phoneme posterior probabilities learned by the acoustic model are then decoded into a word-level representation with $L$ and the grammar information (i.e the language model), $G$. We use a 4-gram language model (LM) using the SRILM toolkit \cite{stolcke2002srilm} trained on the same lyrics corpus with the ones in \cite{demirel2020,dabike2019automatic} which consists of recent English pop songs.

Results are reported on three evaluation sets (see Table \ref{table:data}). The first set is the test split of the \textit{DAMP - Sing! 300x30x2} data set \footnote{The data set is available for research upon request at https://ccrma.stanford.edu/damp.} provided by Dabike et al.\ \cite{dabike2019automatic}. Other evaluation sets are the sung (``\textit{NUS\_sing}") and spoken (``\textit{NUS\_read}') splits of the NUS corpus excluding the native English speakers used in the phoneme analysis in Section 3. For experiments, we have manually segmented the NUS Corpus on the sentence level.

\section{Results}

In the first stage of experiments, we test the benefit of different lexicon extension methods. In Table IV, $L_{CMU}$ denote the standard CMU lexicon. $L_1$ and $L_2$ stand for extended lexicons where alternative pronunciations are generated via removing omitted (low-confidence) consonants and extending vowels (as explained in Section \ref{sec:lexicon_ext}) separately. $L_3$ is then the final singing adapted lexicon which is a combination of both extension methods.

\vspace{5mm} 

\begin{table}[h!]
    \centering
\scalebox{0.82}{
\begin{tabular}{ c !{\vrule width 2pt} c | c | c | c}
  & $L_{CMU}$ & $L_1$ & $L_2$ & $L_3$   \\
  \hline
DAMP\_test & 17.01 & 16.52 & 15.85 & \textbf{15.49} \\
NUS\_read & 9.83 & 9.35 & 9.65 & \textbf{9.40} \\
NUS\_sing & 11.57 & 10.61 & 10.30 & \textbf{9.80} \\
\end{tabular}}
\label{res}
\caption{WERs of different lexicon variants}

\end{table}

These initial results show that proposed lexicon extension methods are overall beneficial for sung word recognition, although $L_1$ resulted in rather more marginal improvement compared to $L_2$. Combining both extension methods achieved the best performance with a relative improvement of $8.24\%$ WER w.r.t to $L_{CMU}$.

Further in Table \ref{table:res2}, we provide the word and character recognition results where the main comparison is between the recognition performances using the standard CMU dictionary and our singing-adapted version ($L_3$), in terms of the error ($\mathit{ER}$), substitution ($S$), insertion ($I$), deletion ($D$) rates explicitly. The singing-adapted dictionary performs consistently better than the speech dictionary even though it can be considered a modest improvement. Note that most of the improvements come from the reduced number of deletions, while the improvement in insertions is generally marginal. 

\begin{table}[ht]
\setlength{\tabcolsep}{0.5pt}
\scalebox{0.82}{
\begin{tabular*}{0.5\textwidth}{
  l 
  S[table-format=-3.4]
  S[table-format=-0.2]
  S[table-format=-0.2]
  S[table-format=2.4] |
  S[table-format=-0.4] 
  S[table-format=-0.4] 
  S[table-format=-0.4]
  S[table-format=-0.4]
  S[table-format=-0.4]
}
& \mc{4}{c}{$L_{CMU}$}   
 & \mc{4}{c}{$L_{3}$}  \\
\cmidrule{2-9}
& {$\mathit{ER}$} &  {$S$} & {$I$} & {$D$} 

& {$\mathit{ER}$} &  {$S$} & {$I$} & {$D$} \\
\cmidrule{2-5} \cmidrule{6-9}
\mc{5}{@{}l|}{\textbf{word}}\\
DAMP\_test &  17.21 &   10.67     &   1.43  &   5.66 &   \textbf{15.49}  & 10.73 & 1.53 & 3.12 \\
NUS\_read &   10.51 &     7.52  &   1.07 &    1.91 &    \textbf{9.40}  & 6.53 & 1.07 & 1.80 \\
NUS\_sing &  13.19  &   8.60    &  1.63 &  2.95 &   \textbf{9.80}  & 6.90 & 1.26 & 1.54 \\
\cmidrule{2-5} \cmidrule{6-9}
\mc{5}{@{}l|}{\textbf{character}}\\
DAMP\_test &   11.41 &  4.78  &   1.85 &  4.79 &  \textbf{9.41}     & 4.25 & 1.63 & 3.53 \\
NUS\_read &   5.57 &  2.73 &   1.38 &    1.47 &   \textbf{5.11}  & 2.54 & 1.11 & 1.36 \\
NUS\_sing &  7.05 & 3.02 &  1.58 &  2.33 &  \textbf{6.14} & 3.03 & 1.36 & 1.75  
\end{tabular*}}
\caption{Word and character error rates using standard (\textit{L\_CMU}) and singing-adapted (\textit{L\_sing}) pronunciation dictionaries.}
\label{table:res2}
\end{table}

\begin{table}[ht]
\centering
\setlength{\tabcolsep}{0.5pt}
\scalebox{0.78}{
\begin{tabular*}{0.5\textwidth}{
  l 
  S[table-format=-3.4]
  S[table-format=-0.2]
  S[table-format=-0.4] |
  S[table-format=2.4] 
  S[table-format=-0.4] 
  S[table-format=-0.4] 
  S[table-format=-0.4]
  S[table-format=-0.4]
}
& \mc{3}{c}{$L_{CMU}$}   
 & \mc{3}{c}{$L_{3}$}  \\
\cmidrule{2-7}
& {$\mathit{ER}$} &  {$S$}  & {$D$} 

& {$\mathit{ER}$} &  {$S$}  & {$D$} \\
\cmidrule{2-4} \cmidrule{5-7}
\mc{4}{@{}l|}{\textbf{word (ending with consonants \textit{D,DH,T,Z})}}\\
DAMP\_test &   22.84 &   13.06  & 7.78  & \textbf{17.67} &  10.15    & 7.21 \\
NUS\_read &   9.74 &       8.82  &    0.91 &    \textbf{9.01}  & 7.90 & 1.10 \\
NUS\_sing &  14.01 &      7.76 &  5.73 &   \textbf{7.94} & 5.73 &  2.21\\
\cmidrule{2-4} \cmidrule{5-7}
\mc{4}{@{}l|}{\textbf{vowel}}\\
DAMP\_test &   13.20 &      6.47  &      6.72 &    \textbf{9.80}  & 5.59 & 4.21 \\
NUS\_read &   4.02 &       2.44 &     1.58 &    \textbf{3.99}  & 2.55  & 1.44 \\
NUS\_sing &  7.23 &      2.98 &    4.26 &  \textbf{ 6.71} & 3.03 &  3.68 \\
\end{tabular*}}
\caption{Error analysis w.r.t omitted consonants and vowels}
\label{res:final}
\end{table}

According to Table \ref{res:final}, $L_{3}$ shows more than absolute 5\% lower ER on singing data. Less words are substituted and deleted using $L_{3}$. The vowel recognition rate is obtained via comparing vowels in the human phoneme annotations and the phonetic transcript of the recognizer. The phonetic transcript is obtained similarly as explained in Section 3. The singing-adapted dictionary also performs consistently better than the speech version with regards to the vowel recognition rate although the improvement in the speech data is marginal, similarly for its response to words ending with low-confidence consonants. This shows that the adaptation is more singing specific, however the improvement is rather modest.

There are further possibilities for adapting the pronunciation model to singing. New alternative pronunciations may be generated via a statistical analysis of the interchange (substitution) of phonemes between speech and singing. Note that these substitutions need to be considered as context-dependent via observing the neighbouring phonemes for the instances of substitution. Additionally, pronunciation probabilities could be extracted from the training data which is reported to be beneficial for word recognition \cite{chen2015pronunciation}. 

\section{Conclusion}

This paper presents a computational approach for an in-depth analysis on the pronunciation differences between singing and speech. The proposed confusion analysis is utilized in identifying systematic pronunciation variances on the phoneme-level. We proposed a new singing-adapted version of the standard CMU dictionary by adding alternative word pronunciations based on the findings of our analysis. We report the best WER scores for ALT from monophonic recordings using an n-gram language model. The error analysis validates our approach being consistently beneficial for sung word recognition. We have publicly shared sentence-level manual annotations on the \textit{NUS Sung and Spoken Lyrics Corpus} to be used as a new benchmark evaluation set for lyrics transcription in monophonic recordings.\footnote{The annotations can be retrieved from https://github.com/emirdemirel/ALTA/s5/data.}

\begin{tikzpicture}[remember picture,overlay]
\node[anchor=south,yshift=10pt] at (current page.south)
{\fbox{\parbox{\dimexpr\textwidth-\fboxsep-\fboxrule\relax}{
\footnotesize  Copyright 2021 IEEE. Published in the IEEE 229th European Signal Processing Conference, EUSIPCO 2021, scheduled for 23-27 August, 2021, in Dublin, Ireland. Personal use of this material is permitted. However, permission to reprint/republish this material for advertising or promotional purposes or for creating new collective works for resale or redistribution to servers or lists, or to reuse any copyrighted component of this work in other works, must be obtained from the IEEE. Contact: Manager, Copyrights and Permissions / IEEE Service Center / 445 Hoes Lane / P.O. Box 1331 / Piscataway, NJ 08855-1331, USA. Telephone: + Intl. 908-562-3966.
}}};
\end{tikzpicture}

\end{document}